# Investigating the Synthetic Minority class Oversampling Technique (SMOTE) on an imbalanced Cardiovascular Disease (CVD) dataset

Ioannis D. Apostolopoulos
School of Medicine
University of Western Greece, Patras, Achaia, Greece
ece7216@upnet.gr

*Abstract*— **In this work, we employ the Synthetic Minority Oversampling Technique (SMOTE) to generate instances of the minority class in an imbalanced Coronary Artery Disease dataset. We firstly analyze the public dataset Z – Alizadeh sani, a dataset used for non-invasive prediction of CAD. We perform feature selection to exclude attributes unrelated to Coronary Artery Disease risk. The generation of new samples is performed using SMOTE, a technique commonly employed in machine learning tasks. We design Artificial Neural Networks, Decision Trees, and Support Vector Machines to classify both the original dataset and the augmented. The results demonstrate that data augmentation may be beneficial in specific cases, but it is not a panacea, and its application in a specific dataset should be carefully examined.**

## I. INTRODUCTION

An imbalanced data set is defined as two-class data set in which one class (called majority) has an overwhelming number of instances than the other class (called minority). The classification problem for imbalanced data is exciting and challenging to researchers because most standard data mining methods perform adequately on balanced datasets but are not applicable for imbalanced ones [1].

The application of Synthetic Minority Oversampling Technique (SMOTE) [2] is widely used in economics, industry, and other fields. It has proven a useful technique to increase the number of data [3]. However, it is not a panacea in all situations. Medical data are a special kind of data; they derive from human beings; therefore, applying oversampling techniques is challenging [4]. In essence, the generation of hypothetical patient cases is risky, and its implementation must be strictly thought.

In this work, we investigate the ability of specific machine learning algorithms to improve their accuracy and reduce their False Positives and False Negatives with augmented medical data. We show that the public dataset named Z-Alizadeh sani [5], is heavily imbalanced.

The prementioned dataset is commonly used for the prediction of CAD [6-10]. The prediction and diagnosis of Coronary Artery Disease (i.e., the stenosis of the coronary arteries of our heart), remains an open issue. Diagnostic tests and symptoms (Angina) are not trustworthy enough for a doctor to directly diagnose CAD [11]. Therefore, the most common practice to confirm or deny the presence of the disease is the invasive Coronary Angiography [12]. Therefore, several machine learning, deep learning, and more generally, artificial intelligence techniques have been proposed to solve the issue [13-16]. The absence of balanced, correctly labeled, and complete datasets is making the task even more challenging.

In this work, we extract specific features from the imbalanced dataset called Z-Alizadeh sani. We then employ SMOTE to generate new instances of the minority class. We design basic architectures of classifiers in order to test their performance on augmented and non-augmented data. Finally, we show that SMOTE can be useful to particular algorithms and cases, as it can improve the classifiers' capability to recognize the False Negatives and False Positives.

The rest of the paper is organized as follows. The imbalance of the dataset and the minority class oversampling technique are explained in section II. Experimental results are presented in section III. Concluding remarks are given in section IV.

## II. DATASET AND METHODS

### A. Minority Class oversampling

The Synthetic Minority Oversampling Technique (SMOTE) was introduced by Chawla et. al (2002) [17]. SMOTE is an oversampling approach for the increase of the minority class instances. The minority class is over-sampled by creating "synthetic" examples rather than by over-sampling with duplicated real data entries. Depending upon the amount of over-sampling required, neighbors from the k nearest neighbors of a record are randomly chosen. Our implementation currently uses five nearest neighbors.

The main process of SMOTE is to find K – nearest neighbors, which defined as the K elements, belong to the minority class for each minority class sample $x_i$ and then randomly selects one of these neighbors. Utilizing interpolation theory, new samples are generated, therefore avoiding the duplication of random instances.

### B. The imbalance of the dataset

From the 59 attributes of the dataset, we use 22 attributes to generate a version of this set. Some attributes do not play an important role in the Coronary Artery Disease diagnosis problem [18], and thus they were excluded from this version. The final attributes are given in Table 1.

Table -1 Extracted Features

| Attribute | Values |
|---|---|
| Typical Angina | 0,1 |
| Atypical Angina | 0,1 |
| Dyspnea | 0,1 |
| Asymptomatic | 0,1 |
| Male | 0,1 |
| Female | 0,1 |
| Age under 40 yrs. | 0,1 |
| Age 40-50 yrs.. | 0,1 |
| Age 50-60 yrs. | 0,1 |
| Age above 60 yrs. | 0,1 |
| Previous Stroke | 0,1 |
| Ex-Smoker | 0,1 |
| Current Smoker | 0,1 |
| Arterial Hypertension | 0,1 |
| Dyslipidemia | 0,1 |
| Obesity | 0,1 |
| Diabetes | 0,1 |
| Chronic Kidney Failure | 0,1 |
| ECG Normal | 0,1 |
| ECG Abnormal | 0,1 |
| ECHO Normal | 0,1 |
| ECHO abnormal | 0,1 |
| CAD/NO CAD | "YES", "NO" |

Figure 1 illustrates the number of Healthy and Diseased instances, proving the strong imbalance of the dataset.

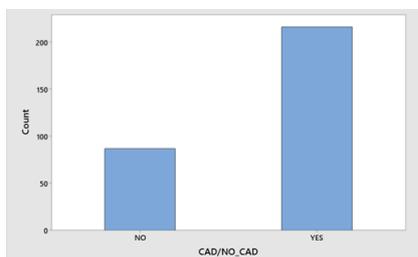

Fig. 1. Dataset's classes population

The number of healthy instances is 87, whereas the number of diseased is 217. Practically that means that if a classifier labels all instances as diseased, it automatically achieves 71.3% accuracy since 71.3% of the instances are, in fact, diseased.

## III. EXPERIMENT AND RESULT

We apply SMOTE using WEKA, and inspect the results of the algorithm. For each instance of the minority class, we generate the nearest neighbors. The final instances of the healthy cases in the augmented dataset are 209. That is, 122 newly generated instances emerged.

The architectures and parameters of ANN (Artificial Neural Network) and trees are given in Table 2. We specify the parameters of the networks we developed, for example, the Learning Rate, the Momentum, the Weight Decay, etc.

Table -2 Parameters of the algorithms

| Classifier / Parameters | Epochs / Batch Size | Hidden Layers / Size | Other |
|---|---|---|---|
| ANN SA1 | 150 / 32 | 512 | LR: 0.1, Mom:0.1, Decay: Yes |
| ANN SA2 | 150 / 32 | 1024 | LR: 0.1, Mom:0.1, Decay: Yes |
| ANN SA3 | 150 / 32 | 1024 | LR: 0.01, Mom:0.1, Decay: Yes |
| Spegasos | 250 / 32 | - | Lambda: 1e-4, Hinge Loss |
| Random Forest | 150 / 32 | Unlimited | - |
| RepTree | Unspecified / 32 | Unlimited | - |

We provide the results in accuracy and incorrectly classified instances of each classifier in each case. The most vital factor and evaluation criteria are the actual reduction of False Positives or False Negatives, and not an improvement in classification accuracy. Besides, an accuracy improvement may, in fact, come from the generated instances (if they are correctly classified) and not from an actual learning improvement of the classifier. Table 3 illustrates the results of the classifiers. For each case, we provide the number of incorrectly classified instances. The experiments were performed using data split. We separate the datasets into the train set that contains 85% of the instances and into the test set that contains 15% of the instances. The test sample size on the original data is 45 instances, and the test sample size coming from the augmented data is 64.

Table -3 Results on original and on augmented data

| Classifier / Metrics | Augmented Data Accuracy | Augmented Data Incorrectly Classified Instances (total) | Augmented Data actual Incorrectly Classified Instances (total) | Original Data Accuracy | Original Data Incorrectly Classified Instances (total) |
|---|---|---|---|---|---|
| ANN SA1 | 82.81 | 11 (64) | 8 (61) | 82.22 | 8 (45) |
| ANN SA2 | 81.25 | 12 (64) | 10 (62) | 82.22 | 8 (45) |
| ANN SA3 | 76.56 | 15 (64) | 10 (59) | 68.8 | 14 (45) |
| Spegasos | 81.25 | 12 (64) | 12 (64) | 77.7 | 10 (45) |
| Random Forest | 89.06 | 7 (64) | 7 (64) | 84.44 | 7 (45) |
| RepTree | 85.9 | 9 (64) | 9 (64) | 75.55 | 11 (45) |

RandomForest classifier outmatches the others, achieving 84.44% accuracy and only seven incorrectly classified instances in the original dataset. Augmentation on the dataset did not decrease the actual mistakes; however, the classifier improved its accuracy (89.06%). In essence, the improvement in accuracy comes from the correct classification of the generated instances and does not depict an actual improvement.

ANN SA3 classifier was benefited from the application of SMOTE. Not only did the accuracy increased from 68.8% to 76.56%, but also the classifier achieved an actual reduction of mistakes. The incorrectly classified instances in the original dataset were 14, whereas the actual incorrectly classified instances in the augmented data were only 10. The classifier classified incorrectly five generated instances.

RepTree was also benefited from data augmentation, reducing its actual mistakes from 11 to 9, and improving its accuracy by +10.45%. Spegasos and ANN SA1 were not benefited from data augmentation, despite their accuracy being slightly increased.

IV. CONCLUSION

The learning capabilities of any classifier depend strongly on the quality of the dataset it utilizes to learn and make predictions. Imbalanced datasets may lead a classifier to learn relations that do not depict the reality, and thus the classifiers are overfitting. We employed SMOTE to generate new instances of the minority class (healthy cases) of Z-Alizadeh sani dataset. We designed six networks and investigated their performance on augmented data, compared to their success rates on the original dataset.

Due to the fact that the results were diverse, we conclude that data augmentation using SMOTE should be deeply examined and carefully though before it is used on medical datasets. Depending on the networks we employ for a specific task, SMOTE can indeed aid to the improvement of mining information from an augmented data, as our experiments demonstrated. However, it is not a panacea for every task, as there were cases in our experiments, where the newly generated instances were not beneficial.